\documentstyle[twocolumn]{article}
\evensidemargin \leftmargin
\makeatletter

\newcounter{subfigure}[figure]

\def\thesubfigure{(\alph{subfigure})\space}

\def\subcapsize{\footnotesize}

\def\subfigtopskip{10pt}

\def\subfigcapskip{10pt}

\def\subfigcapmargin{10pt}

\def\subfigure{%
  \leavevmode
  \@ifnextchar [%
    \@subfigure
    {\@subfigure[\@empty]}}

\long\def\@subfigure[#1]#2{%
  \stepcounter{subfigure}%
  \setbox\@tempboxa \hbox{#2}%
  \@tempdima=\wd\@tempboxa
  \vtop{%
    \vbox{
      \vskip\subfigtopskip
      \box\@tempboxa}
    \vskip\subfigcapskip
    \begingroup
      \@parboxrestore
      \setbox\@tempboxa
      \ifx #1\@empty
        \hbox{\subcapsize\strut\hfil}%
      \else
        \hbox{\subcapsize\strut\thesubfigure#1}%
      \fi
      \@tempdimb=-\subfigcapmargin
      \multiply\@tempdimb\tw@
      \advance\@tempdimb\@tempdima
      \hbox to\@tempdima{%
        \hfil
        \ifdim \wd\@tempboxa >\@tempdimb 
          \parbox{\@tempdimb}{\subcapsize\thesubfigure#1}%
        \else
          \box\@tempboxa
        \fi
        \hfil}
    \endgroup
  \vskip\subfigtopskip}}
\def\subfigtopskip{5pt}       
\def\subfigcapskip{5pt}       
\def\subfigcapmargin{5pt}     
\input{epsfig.sty}      
\newcount\@tempcntc
\def\@citex[#1]#2{\if@filesw\immediate\write\@auxout{\string\citation{#2}}\fi
  \@tempcnta\z@\@tempcntb\m@ne\def\@citea{}\@cite{\@for\@citeb:=#2\do
    {\@ifundefined
       {b@\@citeb}{\@citeo\@tempcntb\m@ne\@citea\def\@citea{,}{\bf ?}\@warning
       {Citation `\@citeb' on page \thepage \space undefined}}%
    {\setbox\z@\hbox{\global\@tempcntc0\csname b@\@citeb\endcsname\relax}%
     \ifnum\@tempcntc=\z@ \@citeo\@tempcntb\m@ne
       \@citea\def\@citea{,}\hbox{\csname b@\@citeb\endcsname}%
     \else
      \advance\@tempcntb\@ne
      \ifnum\@tempcntb=\@tempcntc
      \else\advance\@tempcntb\m@ne\@citeo
      \@tempcnta\@tempcntc\@tempcntb\@tempcntc\fi\fi}}\@citeo}{#1}}
\def\@citeo{\ifnum\@tempcnta>\@tempcntb\else\@citea\def\@citea{,}%
  \ifnum\@tempcnta=\@tempcntb\the\@tempcnta\else
   {\advance\@tempcnta\@ne\ifnum\@tempcnta=\@tempcntb \else \def\@citea{--}\fi
    \advance\@tempcnta\m@ne\the\@tempcnta\@citea\the\@tempcntb}\fi\fi}
\def\thebibliography#1{\section*{\refname}
    \@mkboth{\uppercase\refname}{\uppercase\refname}
    \raggedright\small
    \list{\@biblabel{\arabic{enumiv}}}
       {\settowidth\labelwidth{\@biblabel{#1}}%
        \parsep  0pt plus 1pt minus 1pt
        \itemsep 0pt plus 0pt minus 0pt
        \itemindent -0.7em
        \listparindent \itemindent
        \leftmargin-\itemindent
        \advance\leftmargin\labelwidth \advance\leftmargin\labelsep
        \usecounter{enumiv} \let\p@enumiv\@empty
        \def\theenumiv{\arabic{enumiv}}}%
    \def\newblock{\hskip .11em plus.33em minus.07em}%
    \sloppy\clubpenalty4000\widowpenalty4000
    \sfcode`\.=\@m}
\@namedef{array*}#1{\def\@halignto{to#1}\@nnarray}
\def\@nnarray{\let\@acol\@arrayacol \let\@classz\@arrayclassz
 \let\@classiv\@arrayclassiv
 \let\\\@arraycr\@tabarray}
\expandafter \let \csname endarray*\endcsname = \endarray
\makeatother
\newcommand{\mmath}[1]{{\ifmmode{#1}\else $#1$\fi}}
\newcommand{\text}[1]{{\mbox{#1}}}
\newcommand{\txt}[1]{{\mbox{#1}}}
\newcommand{\PKzS}{{\mmath{\mathrm{K^0_S}}}}
\newcommand{\PZz}{\mmath{\mathrm{Z^0}}}
\newcommand{\Pp}{\mmath{\mathrm{p}}}
\newcommand{\MSbar}{{\mmath{\overline{\mr{MS}}}}}
\newcommand{\lameff}{{\mmath{\Lambda_{\txt{eff}}}}}
\newcommand{\lqcd}{{\mmath{\Lambda_{\txt{QCD}}}}}
\newcommand{\qh}{{\mmath{Q^\prime_0}}}

\newcommand{\ebeam}{\mmath{E_{\txt{beam}}}}
\newcommand{\beq}{\begin{equation}}
\newcommand{\eeq}{\end{equation}}
\newcommand{\bea}{\begin{eqnarray}}
\newcommand{\eea}{\end{eqnarray}}

\newcommand{\mr}[1]{\mathrm{#1}}
\newcommand{\order}{{\cal{O}}}

\newcommand{\mcol}{\multicolumn}

\newcommand{\GeV}{\txt{G$e$V}}
\newcommand{\MeV}{\txt{M$e$V}}

\newcommand{\as}{{\alpha_s}}
\newcommand{\fsc}[1]{\mbox{\sc #1}}
\newcommand{\thalf}{{\textstyle\frac{1}{2}}}
\begin{document}
\title{\thispagestyle{empty}\vskip 2.5cm
 \hfill \parbox{6cm}{\Large
                       {\bf IC/HEP/95-2} \\
                       to be published in \\ Zeit. f\"ur Physik C}
 \vskip 1.5cm
   \parbox{\hsize}{\centering\bf\Huge
   Experimental  momentum spectra of identified hadrons
   at $e^+e^-$ colliders compared to QCD calculations
  }
 \vskip 1.5cm
}
\author{\parbox{\hsize}{\centering\LARGE
   N.C.~Br\"ummer   \\
   Imperial College London \\
   e-mail: N.Brummer@ic.ac.uk \\
 \vspace{1cm}
}}
\date{}
\thispagestyle{empty}
\begin{titlepage}
\thispagestyle{empty}
\maketitle
\normalsize
\begin{abstract}
Experimental data on the shape of hadronic momentum spectra
are compared to theoretical predictions in the context of
calculations in the Modified Leading Log Approximation (MLLA),
under the assumption of Local Parton Hadron Duality (LPHD).

Considered are experimental measurements at $\mathrm{e^+e^-}$-colliders
of $\xi_p^*$, the position of the maximum in the distribution of
$\xi_p=\log(1/x_p)$, where $x_p=p/p_{\text{beam}}$.
The parameter $\xi_p^*$ is determined
for various hadrons at various centre of mass energies.
The dependence on the hadron type
poses some interesting questions about the process of hadron-formation.
The dependence of $\xi^*_p$ on the centre of mass energy is
seen to be described adequately by perturbation theory.
A quantitative check of LPHD + MLLA is possible by extracting a value
of $\alpha_s$ from an overall fit to the scaling behaviour of~$\xi^*_p$.
\end{abstract}
\end{titlepage}
\section{Introduction}

During a few years of LEP running,
a large amount of information has been collected on iden\-ti\-fied
had\-ron spe\-cies in jets.
The higher centre of mass energy of LEP, compared to past $\mr{e^+e^-}$
accelerators, makes it easier to separate the behaviour
of hadrons with a high momentum,
which are correlated strongly to the primary quark,
from those with a low momentum, created mainly
during fragmentation and hadronisation.

Clear scaling violations have been observed in the
shape of the charged particle distribution of $x_p=p/E_{\text{beam}}$
as a function of the centre of mass energy.
The strong coupling constant has been extracted from
these scaling violations, using the behaviour at
high momenta: $0.2 < x_p < 0.7$ \cite{deboer+kuszmaul}.
Due to the larger statistical errors this is not possible
for individually identified hadron species.

\begin{figure}[btp]
\begin{center}
\mbox{\epsfig{file=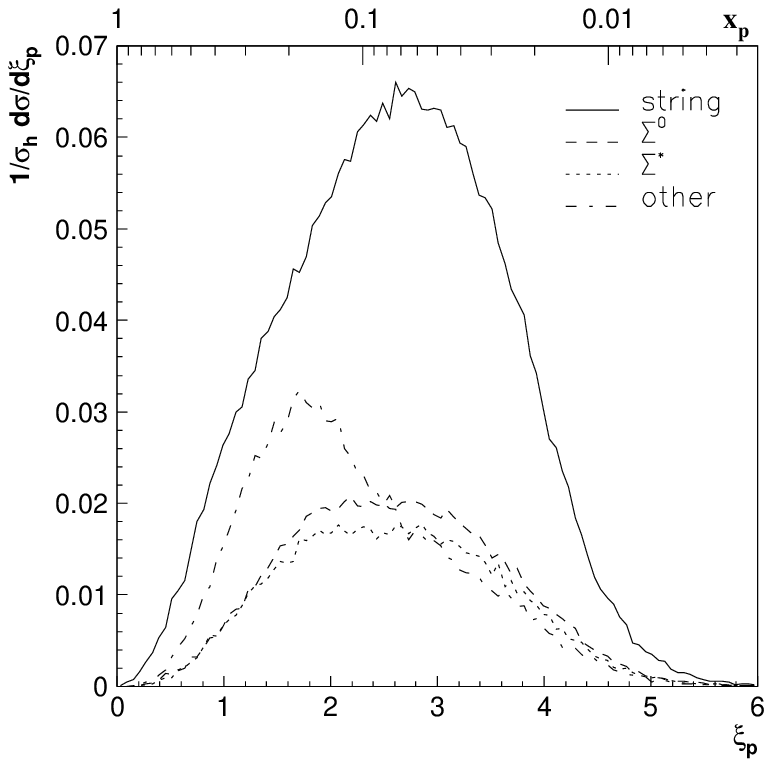,width=.98\hsize}}
\\
\caption{\label{f:xp-lambda-father}
The dependence of the $\Lambda$ spectrum on
the species of its `parent' particle in
$\PZz$ decays simulated by {\tt JETSET}.
Particles produced directly in the fragmentation
process have the Lund-model string as `parent'.
The scales of both $x_p$ and $\xi_p$ are given for comparison.
}
\mbox{\epsfig{file=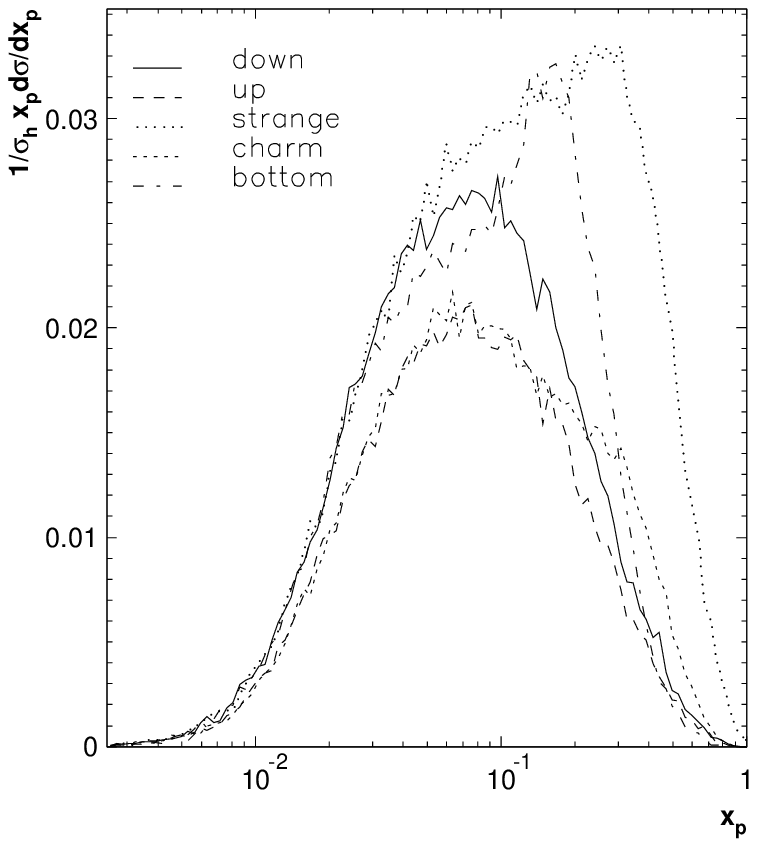,width=.98\hsize}}
\\
\caption{\label{f:xp-lambda-flavour}
The dependence
of the $\Lambda$ spectrum on the flavour  of the primary quarks
in \PZz\ decays at LEP, as predicted by {\tt JETSET}.
Note that there is not a clear separation into a region of high
momentum (correlated to primary quarks)
and another of low momentum (related to jet fragmentation).
}
\end{center}
\end{figure}
\begin{table*}[tb]
\centering
{\footnotesize
\(\begin{array*}{\textwidth}
{@{\extracolsep{\fill}}|l
@{\hspace{1pt}}r@{\extracolsep{0pt}}l
@{\hspace{2pt}\extracolsep{\fill}}
*{8}{@{\hspace{2pt}}c}|}
\hline
& \mcol{2}{c}{\sqrt{s}}
& \mcol{1}{c}{\text{charged}}
& \mcol{1}{c}{\pi^\pm}
& \mcol{1}{c}{\pi^0}
& \mcol{1}{c}{\mr{K}^\pm}
& \mcol{1}{c}{\mr{K}^0}
& \mcol{1}{c}{\mr{p}}
& \mcol{1}{c}{\Lambda}
& \mcol{1}{c|}{\Xi^-\rule[-1.5ex]{0ex}{4ex}} \\
\hline\rule[0ex]{0pt}{3ex}
 \fsc{ARGUS}
             &  9&.98  && 2.32\pm0.022 & 2.42\pm0.08
                        & 1.64\pm0.05 & 1.72\pm0.04
                        & 1.63\pm0.07 & 1.44\pm0.034 & 1.32\pm0.11 \\
 \fsc{CLEO}
             & 10&.49  && 2.30\pm0.08 &
                        & & 1.68\pm0.17
                        &1.67\pm0.10 & 1.47\pm0.08  & \\
\hline\rule[0ex]{0pt}{3ex}
 \fsc{TASSO}
             & 14&      & 2.45\pm0.05
                        & 2.66\pm0.06 &
                        & 1.91\pm0.13 & 1.65\pm0.13
                        & 1.80\pm0.15 & & \\
 \fsc{JADE}
             & 14&      && & 2.68\pm0.27 & & & & & \\
\hline\rule[0ex]{0pt}{3ex}
 \fsc{TASSO}
             & 22&      & 2.74\pm0.06
                        & 2.99\pm0.09 & & 2.41\pm0.23 & 2.47\pm0.8
                        & 2.14\pm0.27 & 1.75\pm0.50 & \\
 \fsc{JADE}
             & 22&.5   & & & 2.84\pm0.30 & & & & &  \\
\hline\rule[0ex]{0pt}{3ex}
 \fsc{HRS}
             & 29&     & & >3.3 &
                        & 2.28\pm0.42 & 2.25\pm0.30
                        & & 1.96\pm0.12 & \\
 \fsc{TPC}/2\gamma
             & 29& & & 3.00\pm0.05 &
                        & 2.14\pm0.05 & 1.98\pm0.09
                        & 2.13\pm0.11 & & \\
\hline\rule[0ex]{0pt}{3ex}
 \fsc{TASSO}
             & 34&      & 3.00\pm0.05
                        & 3.17\pm0.05 & & 2.35\pm0.20 & 2.32\pm0.10
                        & 2.24\pm0.09 & 2.15\pm0.15 &  \\
 \fsc{JADE}
             & 35&      & & & 3.24\pm0.05 & & & & &  \\
\hline\rule[0ex]{0pt}{3ex}
 \fsc{TASSO}
             & 44&      & & 3.37\pm0.07 & & {<}3.5 & & {<}3.5 & & \\
 \fsc{JADE}
             & 44&      & & & 3.52\pm0.08 & & & & & \\
\hline\rule[0ex]{0pt}{3ex}
 \fsc{TOPAZ}
             & 58&      & 3.42\pm0.04
                        & 3.51\pm0.07 & & 2.83\pm0.10 & & 2.64\pm0.06 & & \\
\hline\rule[0ex]{0pt}{3ex}
 \fsc{OPAL}
             & 91&.2    & 3.60\pm0.04
                        & 3.79\pm0.04& &2.73\pm0.06& 2.87\pm0.07 &2.95\pm0.09
                        & 2.78\pm0.08 & 2.55\pm0.20 \\
 \fsc{DELPHI}
             & 91&.2    & 
                        & & & & 2.62\pm0.11 & & 2.81 \pm 0.04 & \\
 \fsc{L3}
             & 91&.2    &3.71\pm0.05
                        & & 3.96\pm0.13 & &2.89\pm0.05 & & & \\
\hline
\end{array*}\)}
\\
\caption{The values of $\xi^*_p$ for various hadrons
as determined from the momentum spectra as measured by
$\mr{e^+e^-}$ experiments at different values of $\protect\sqrt{s}$.
For the $\eta$ meson, the L3 experiment reported
a value $\xi^*_p=2.60\pm0.15$.
\label{t:expxist}}
\end{table*}

However, low momentum data for specific types of hadrons may be used
to study the properties of
jet-evolution and hadron-formation in the context of the LPHD hypothesis
and MLLA calculations of parton spectra
\cite{book-ptqcd,mlla-xp-mp,mlla-xp-Leff,mlla-xp}.
The assumption of `Local Parton Hadron Duality' (LPHD) states
that a calculated spectrum for `partons' in a `parton shower' can be
related to the spectrum of real hadrons by simple normalisation constants.
These constants have to be determined by experiment.
A second assumption
is that the low momentum part of the spectrum is not influenced in
a significant way by hadrons that are correlated to the primary quark.

Calculations of the parton spectra
in the `Modified Leading Log Approximation' (MLLA)
take into account next-to-leading logarithms in a consistent fashion.
The physical mechanism relevant to these next-to-leading
logarithms is the coherent emission of soft gluons
inside a jet, leading to an angular ordering and an effective
transverse momentum cutoff for the partons.
Parton jets develop through repeated parton splittings, resulting
in an increase of the multiplicity at lower momenta.
The interplay of coherent emission of gluons and the creation of
hadrons causes this spectrum to be cut off at very low momenta.
Calculations predict the shape of the distribution of $\xi_p=\log(1/x_p)$.
The resulting `hump-backed' distribution is nearly gaussian.
As an example, the Monte Carlo spectrum of
the $\Lambda$ baryon in \PZz\ decays can be seen in
figures \ref{f:xp-lambda-father} and \ref{f:xp-lambda-flavour}.
In the following, the maximum $\xi^*_p$ in the $\xi_p$ distribution will be
determined for various types of hadrons and at various centre
of mass energies. Subsequently a comparison is made with
the theoretical calculations.

Figure~\ref{f:xp-lambda-father} shows that LPHD is not an obvious
assumption: in the {\tt JETSET} Monte Carlo \cite{JETSET},
the spectrum of e.g.\ a $\Lambda$ baryon depends on its
`parent' particle. The LPHD assumption states that
the sum of these spectra is proportional to the
spectrum as calculated for a parton shower
with a correctly chosen energy cut-off.

Moreover, the {\tt JETSET} momentum spectrum of the $\Lambda$ baryon
(figure~\ref{f:xp-lambda-flavour}) shows
a strong dependence on the flavour of the primary quarks.
For the heavier quarks, $\mr{s}$, $\mr{c}$ and $\mr{b}$,
the flavour dependence does not only manifest itself at high
momenta, but also affects the distribution at low momenta.
This will cause shifts of $\xi^*_p$ that depend on the jet-flavour.

The spectrum of all charged particles at LEP
has been fitted to the MLLA distribution,
with free normalisation factors for pions, kaons and protons,
depending on the centre of mass energy \cite{OPAL_coherence,L3_pi0}.
It was quite surprising \cite{mlla-xp-Leff}
that the MLLA functions can also fit
the high momentum part in the data for pions.
This is probably a coincidence, since
the heavier $\mr{K^0}$ meson has a spectrum
that can not be fitted as nicely at large $x_p$.

\begin{figure*}[bt]
\makebox[\hsize]{
\subfigure[$\xi^*_p$ versus the hadron mass]
          {\mbox{\epsfig{file=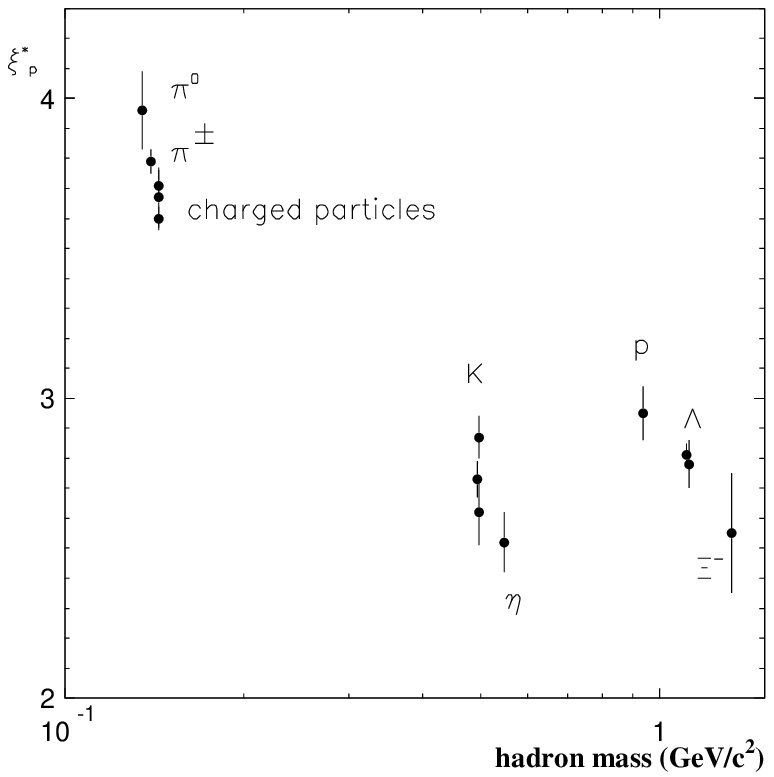,width=.33\hsize}}}
\hfill
\subfigure[$Q_0$ versus the hadron mass]
          {\mbox{\epsfig{file=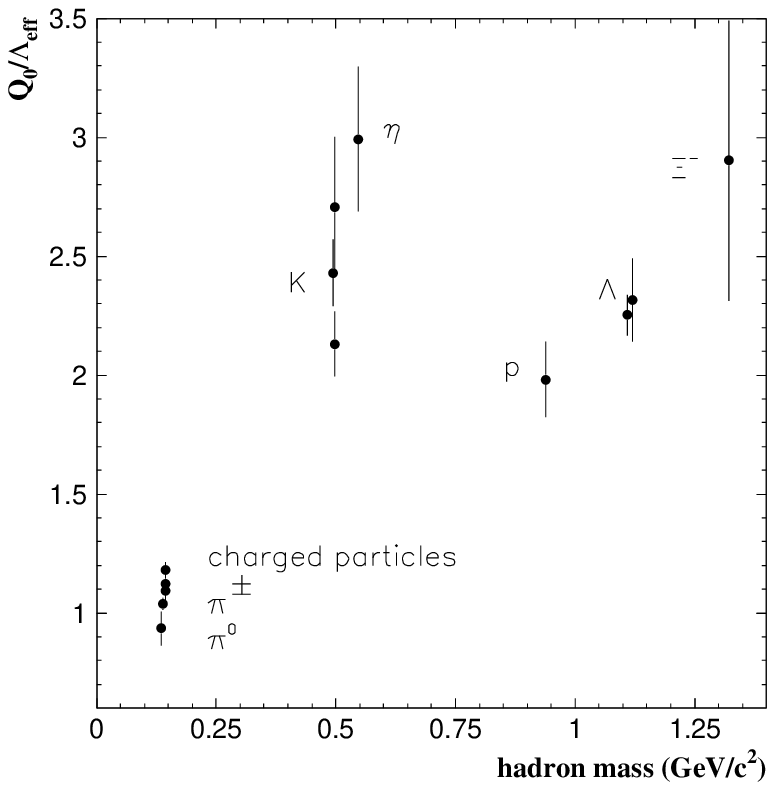,width=.33\hsize}}}
\hfill
\subfigure[$Q_0$ from fit versus the hadron mass per quark]
          {\mbox{\epsfig{file=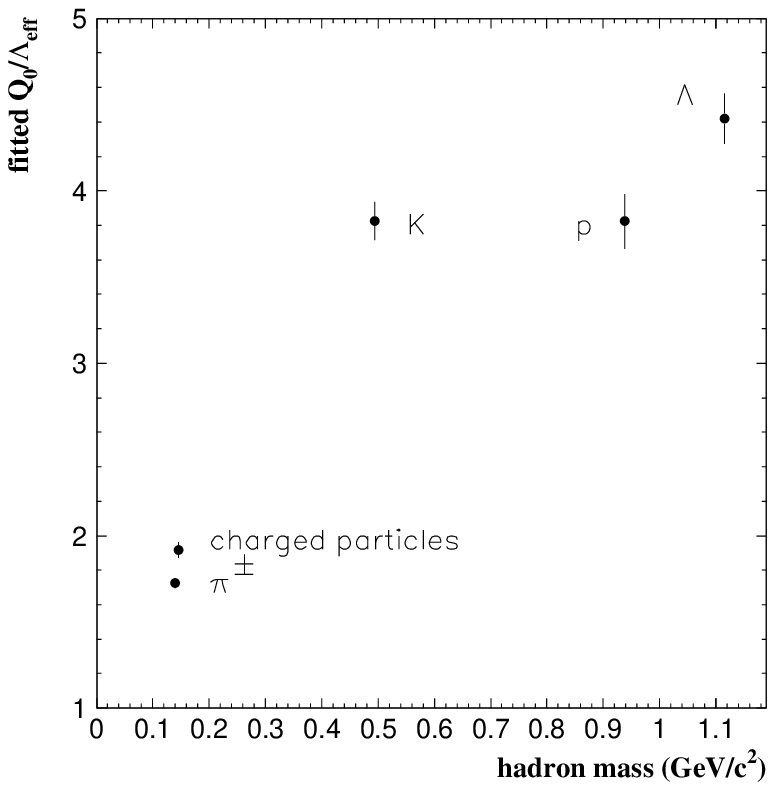,width=.33\hsize}}}
}
\par
\caption{The maxima $\xi^*_p$ of the distributions
in $\xi=-\log x_p$ for different hadrons as measured at LEP
(table \ref{t:expxist}) are shown in figure (a).
Two different determinations of the corresponding scale $Q_0$ are
shown in figures (b) and (c).
The calculated dependence for $\Lambda_{\protect\text{eff}} = 50\ \MeV$
leads to figure (b), where one would naively expect
$Q_0$ to be proportional to the hadron mass.
Figure (c) gives the values of $Q_0$ as determined from
the fit of the dependence of $\xi^*_p$ on $\ebeam$.
A (large) common error on the different values of
$Q_0$ has been neglected in both figures (b) and (c).
This causes the vertical scales to be rather arbitrary, but the
relative position of the points remains significant.
\label{f:ximax}}
\end{figure*}

One way to see whether the observed shape of the spectrum is really due to
coherent gluon emission is to compare experimental
data to predictions of the {\tt JETSET} Monte Carlo program,
with the coherence (angular ordering) either turned on or turned off.
The large dependence on the primary quark flavours and the
fact that coherence in {\tt JETSET} is not really necessary
to fit the experimental data
has led some to the conclusion \cite{no-coherence-evidence}
that coherence can not be demonstrated on the basis of hadron
momentum spectra.

Here an attempt will be made to infer more information
using the properties of identified hadrons, concentrating
on the observable $\xi^*_p$.
This is done for data from LEP and from various other $\mr{e^+e^-}$ colliders.
Subsequently, the available results for $\xi^*_p$ are compared to the
LPHD + MLLA approach \cite{book-ptqcd,mlla-xp-mp}.


\section{Experimental values of $\xi^*_p$ for identified hadrons}

At LEP,
analyses of explicitly identified hadrons have been performed for
the $\pi^\pm$ \cite{OPAL-dedx},
the $\pi^0$ \cite{L3_pi0,L3-neutrals},
the $\mr{K}^\pm$ \cite{OPAL-dedx},
the \PKzS\ \cite{L3-neutrals,DELPHI_K0L0Xi,OPAL_K0},
the $\eta$ \cite{L3_eta,L3-neutrals}
the proton \cite{OPAL-dedx}
and the $\Lambda$
\cite{DELPHI_K0L0Xi,OPAL_baryon,NB_PhD,DELPHI-LL-corr}.
At lower statistics data is also available
for the $\Xi^-$ and $\Omega^-$ baryons,
as well as the $\Sigma^*$ and $\Xi^*$ \cite{DELPHI_K0L0Xi,OPAL_baryon}.

The values of $\xi^*_p$ correspond to low momenta, where the dependence on
the pri\-ma\-ry quark fla\-vours is ex\-pec\-ted to be small.
To first approximation the distribution in $\xi_p$ is gaussian, but a
distorted gaussian fits the calculated spectrum
more accurately \cite{dgaus,mlla-xp-mp}.
The statistics of identified hadrons is not
sufficient to extract the width and the two distortion parameters for
more than a few experiments.

However, since these other parameters depend more strongly
on the high momentum tail of the distribution,
they can be expected to depend more strongly on the event flavour,
which could bias the results. We will suppose that this violation
of LPHD can be avoided adequately by concentrating on $\xi^*_p$, a
property of the low-momentum part of the spectrum.

In this paper $\xi^*_p$ is simply defined as the maximum of the distribution.
The values were determined
by fitting the $\xi_p$ distribution in a limited range ($\approx \pm 1$)
of $\xi_p$ around the maximum, to a gaussian distribution.
A systematic error was estimated by changing the fitted range and
by fitting the same range to a polynomial function.
The statistical and systematic errors were added in quadrature.
The systematic error often dominated.

For the most accurately determined experimental spectra,
those of identified charged hadrons at OPAL \cite{OPAL-dedx},
a more complicated function is necessary.
In this case a polynomial of order $4$ or $5$ was used to fit
either the partial cross section,
or its logarithm\footnote{Fitting the logarithm to a $4$th order
polynomial is equivalent to using a distorted gaussian.}.
These polynomials were defined such that the maximum was one of the
free parameters in the fit. In this way the MINOS procedure of
MINUIT \cite{MINUIT} can automatically
determine an interpolation error due to variations of the other
free parameters of the polynomial. The values found for the
maxima in OPAL's spectra are different from those reported by OPAL, and
the estimated errors are slightly larger.

The results are given in table~\ref{t:expxist}.
Use was made of spectra published by
CLEO \cite{CLEO-spectra},
ARGUS \cite{ARGUS-spectra},
TPC \cite{TPC-spectra},
HRS \cite{HRS-spectra},
TASSO \cite{TASSO-spectra},
JADE \cite{JADE-spectra},
TOPAZ \cite{TOPAZ-spectra},
DELPHI \cite{DELPHI_K0L0Xi,DELPHI-LL-corr},
OPAL \cite{OPAL_coherence,OPAL_K0,OPAL_baryon,OPAL-dedx}
and
L3 \cite{L3_pi0,L3_eta,L3-neutrals}.
Most of the recent publications reported values of $\xi^*_p$, but they
did not all use exactly the same definition of it.
While the numbers given by L3 were compatible with the method used here,
OPAL and DELPHI used slightly different definitions and for consistency
the values of  $\xi^*_p$ are those determined by the author
from the published experimental spectra.
The spectra from TOPAZ were not available in \cite{TOPAZ-spectra}, only
values of $\xi^*_p$; likewise for the recent paper by L3 \cite{L3-neutrals}.
For most experiments at lower centre of mass energies
the authors did not report the distribution of $\xi_p$ or $x_p$, but used
the energy to define $x_E=2E/E_{\text{beam}}$,
or they divided by $\beta=v/c$ to obtain the `scaling cross section'.
Many different normalisations have been used,
but fortunately this is unimportant for a determination $\xi^*_p$.
 From the experiments (ARGUS and CLEO)
near the $\Upsilon$ resonances results are given for the
`continuum', and for the resonances itself, where
the contributions from the continuum are subtracted.
The values in table~\ref{t:expxist} refer to data from
the continuum above the $\Upsilon$ resonance.
In the data from ARGUS for charged pions and protons, the
contributions from decays of $\mr{K^0_S}$\ and $\Lambda$\ were subtracted.

%
%

\begin{figure}[tb]
\centering
\makebox[\hsize]{\epsfig{file=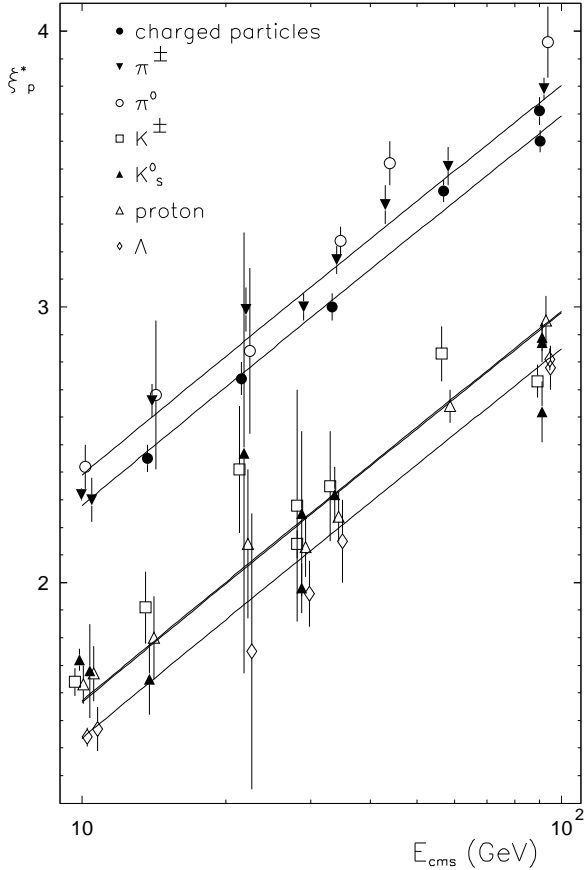,width=\hsize}}
\\
\caption{The experimental values of $\xi_p^*$ versus $\protect\sqrt{s}$
for various identified hadrons.
The curves are from the combined fit
of $\Lambda_{\txt{eff}}$ and $F(\txt{hadron})$ to the predicted
dependence (equations (\ref{eq:xistq2}) and (\ref{eq:fitresults})).
 From top to bottom,
the curves correspond to pions ($\pi^\pm$, $\pi^0$),
all charged particles, kaons ($\mr{K^\pm}$, $\mr{K^0}$),
protons and $\Lambda$ baryons. The curves for kaons and protons happen to
lie on top of each other.
\label{f:xistq2all}}
\end{figure}

\section{Predictions for $\xi^*_p$}

The LPHD + MLLA calculations \cite{mlla-xp-mp}
of the distribution of $\xi_p$
depend on an effective QCD scale $\lqcd\sim\lameff$
and on a transverse momentum cutoff $Q_0$ in the evolution
of the parton cascade.
The value of $\xi^*_p$ is calculated to be a nearly linear
function of $\log Q_0$, $\log\lameff$ and $\log\ebeam$.

The calculated dependence on
the centre of mass energy is as follows \cite{mlla-xp-mp}:
\bea
\xi^*_p &=& Y\left[\frac{1}{2}+\sqrt{C/Y}-C/Y+{\cal O}(Y^{-3/2})\right]
\nonumber \\
         & & {}+ F(\lambda),
\label{eq:xistq2}
\eea
where $F(0)=0$ and
\footnote{This detail \cite{private-yuri} is not completely clear in
reference \cite{mlla-xp-mp}, where equation (\ref{eq:xistq2}) is given
as a function of $Y-\lambda$, but in the approximation that $\lambda=0$.}
\beq
Y=\log(E_{\text{beam}}/\Lambda_{\text{eff}}), \qquad
\lambda=\log(Q_0/\Lambda_{\text{eff}}),
\eeq
and $\Lambda_{\text{eff}}$ is an effective QCD scale, while only
the cutoff scale $Q_0$ depends on the hadron type.
The constant $C$ is calculated to be
\beq
 C=\left(\frac{a}{4N_c}\right)^2 \frac{N_c}{b},
\eeq
where
\bea
a &=& 11N_c/3 +2n_f/(3N_c^2),\label{eq:a} \\
b &=& 11N_c/3-2n_f/3.  \label{eq:b}
\eea
$N_c=3$ is the number of
colours and $n_f=3$ is the active number of quark flavours in the
fragmentation process. For $n_f=3$ one finds $C=0.2915$.

It is important to note that $F$ does not depend on $Y$, and that
the first part of equation (\ref{eq:xistq2}) is independent of $Q_0$.
This predicted behaviour can be checked by comparing
spectra of different identified particles.
In the available momentum range this leads to
a nearly linear dependence of $\xi^*_p$ on $Y$.

In ref.\ \cite{mlla-xp-mp} various graphs show the results
of numerical calculations for the dependence of various
parameters on $E_{\txt{beam}}$ and $Q_0\neq\lameff$.
Figure 4 of ref.\ \cite{mlla-xp-mp} shows the dependence
of $\xi^*_p$ on $Q_0$ for $\lameff=150~\MeV$,
and at a number of beam energies. From this it is
possible to extract $F$ after subtracting the $Y$
dependent part of equation (\ref{eq:xistq2}).
For an interpretation of the experimental values of $\xi^*_p$
it was useful to fit both this function
$F(\lambda)$, and its inverse $\lambda(F)$ to polynomials.
The result for $F(\lambda)$ is:
\beq
  F(\lambda) =
-1.46\cdot\lambda+0.207\cdot\lambda^2\pm0.06 ~~~~
\label{eq:Flambda}
\eeq
while the inverse of $F$ is described by:
\beq
 \lambda(F) =
-0.614\cdot F+0.153\cdot F^2 \pm 0.06
\label{eq:lambdaF}
\eeq
The given errors denote the maximum deviation from the
distribution plotted in ref.\ \cite{mlla-xp-mp}, in the ranges
$-2< F< 0$ and $0<\lambda< 2$.


\section{The meaning of $\Lambda_{\txt{eff}}$}

The QCD scale $\lameff$ \cite{private-yuri}
is related to the (running) strong coupling constant by:
\beq
\frac{\as}{2\pi}= \frac{1}{b\,Y}
                = \frac{1}{b\log(\ebeam/\lameff)}.
\label{eq:asrun}
\eeq
This makes it possible to express the scale dependence of
eq.~(\ref{eq:xistq2}) in terms of $\alpha_s$:
\bea
\frac{\partial\xi^*_p}{\partial Y} =
 \frac12 + \frac18 a\sqrt{\frac{\as}{2\pi N_c}}
  + \order\left(\as^{5/2}\right),
\label{eq:yuri} \\
\frac{\partial^2\xi^*_p}{\partial Y^2} =
\frac{-ab}{16\sqrt{N_c}}\left(\frac{\alpha_s}{2\pi}\right)^{3/2}
  + \order\left(\as^{7/2}\right).
\label{eq:yuri2}
\eea
The $\sqrt{\alpha_s}$
term is the next-to-leading (MLLA) correction to the leading term.
The next-to-next-to-leading term $\order(\alpha_s)$
vanishes \cite{private-yuri}.

Although this sounds impressive, it is still not enough for
an unambiguous definition of $\lameff$ or, equivalently, of
$\alpha_s(E)$ at a well-defined energy scale $E$.
An uncertainty of e.g.\ a factor $2$ in the energy scale leads to
$\alpha_s(2E)$ instead of $\alpha_s(E)$.
This induces a correction in equation (\ref{eq:yuri})
of $\order(\alpha_s^{3/2}(E))$,
which shows that for a better definition of the energy
scale it is necessary to know the $\order(\alpha_s^{3/2}(E))$ correction.
This is the reason for talking about the scale $\lameff$, in stead
of a more well-defined scale like $\Lambda_{\MSbar}$.

If we now neglect the dependence of $\alpha_s(E)$ on $E$, $\xi^*_p$
becomes a linear function of $\log(\ebeam)$, where the slope is
given by equation~(\ref{eq:yuri}).

\begin{table*}[tb]
\centering
{\small $\begin{array*}{\textwidth}{
@{\extracolsep{\fill}}
| l
| r @{\hspace{.2em}}
  r @{\hspace{.3em}}
  r @{\hspace{1.8em}}
| r @{\hspace{.3em}}
  r @{\hspace{1.5em}}
| r @{\hspace{2.5em}}
|}
\hline
   \rule[0pt]{1em}{0em}\rule[-1ex]{0pt}{3.5ex}
   \text{particle}
 & \mcol{1}{ c }{a_h}
 & \mcol{1}{ c }{b_h}
 & \mcol{1}{ c|}{\chi^2/\txt{n.d.f.}}
 & \mcol{1}{ c }{\qh}
 & \mcol{1}{ c|}{\chi^2/\txt{n.d.f.} }
 & \mcol{1}{ c|}{\txt{points}\rule{1em}{0em}}
\\
\hline
\rule[0pt]{1em}{0em}
\text{charged}& 0.76\pm0.12   & 0.641\pm0.030   & 1.36 & 0.212\pm0.007
           & 1.10  & 6 \\
\hline \rule[2.5ex]{1em}{0ex}
\pi^\pm      & 0.79 \pm 0.06  & 0.667 \pm 0.018 & 1.23 & 0.179\pm 0.005
           & 1.53  & 9 \\
\rule[0pt]{1em}{0em}
\pi^0        & 0.78 \pm 0.20  & 0.705\pm 0.059  & 0.42 & 0.147\pm 0.008
           & 0.70  & 6  \\
\rule[0pt]{1em}{0em}
\pi^\pm,\ \pi^0
\rule[0pt]{1em}{0em}
             & 0.78 \pm 0.05  & 0.679\pm 0.016  & 1.49 & 0.174\pm 0.004
           & 1.91  &15  \\
\hline \rule[2.5ex]{1em}{0ex}
\mr{K^\pm}   & 0.47 \pm 0.11  & 0.512\pm 0.032  & 2.17 & 0.645\pm0.027
           & 4.48  &8   \\
\rule[0pt]{1em}{0em}
\mr{K^0}     & 0.49\pm 0.09   & 0.517\pm 0.025  & 2.12 & 0.620\pm0.023
           & 5.49  &10   \\
\rule[0pt]{1em}{0em}
\mr{K^\pm}, \mr{K^0}
             & 0.48 \pm 0.06  & 0.516\pm 0.019  & 1.93 & 0.631\pm0.018
           & 4.78  &18  \\
\hline \rule[2.5ex]{1em}{0ex}
\Pp          & 0.27 \pm 0.12  & 0.580\pm 0.036  & 0.37 & 0.625\pm0.030
           & 0.93  &8   \\
\rule[0pt]{1em}{0em}
\Lambda      & 0.02 \pm 0.07  & 0.613\pm 0.021  & 0.30 & 0.759\pm0.025
           & 1.07  &7   \\
\rule[0pt]{1em}{0em}
{\Pp},\ \Lambda & 0.09 \pm 0.07  & 0.609\pm 0.019  & 1.29 & 0.712\pm0.020
           & 1.69  &15  \\
\hline \rule[2.5ex]{1em}{0ex}
\txt{total}~\chi^2/\txt{n.d.f.} &&              &1.21  &
           & 2.41 & 54  \\
\hline
\end{array*}$}
\newline
\\
\caption{Result of a fit of the experimental values of\ $\xi^*_p$ in
table \ref{t:expxist}
to a linear function $\xi^*_p=a_h+b_h\cdot\log (E_{cms}/\GeV)$\
for various identified hadrons.
The scale\ $\qh$ is the result
of a second fit to the function\
$\xi^*_p=\frac{1}{2}Y+\protect\sqrt{CY}-C$,
where\ $Y\equiv \log(E_{\txt{beam}}/\qh)$ and\ $C=0.2915$.
The fits improve slightly when data at the $\Upsilon$ are excluded.
To compare the overall quality of the fits to the two functions, a total
$\chi^2/\txt{n.d.f.}$ is given, based on the fits for
$\pi^\pm$, $\pi^0$, $\mr{K^\pm}$, $\mr{K^0}$, $\mr{p}$, and $\Lambda$.
\label{t:xistq2}}
\end{table*}
\begin{figure}[tb]
\centering
          {\mbox{\epsfig{file=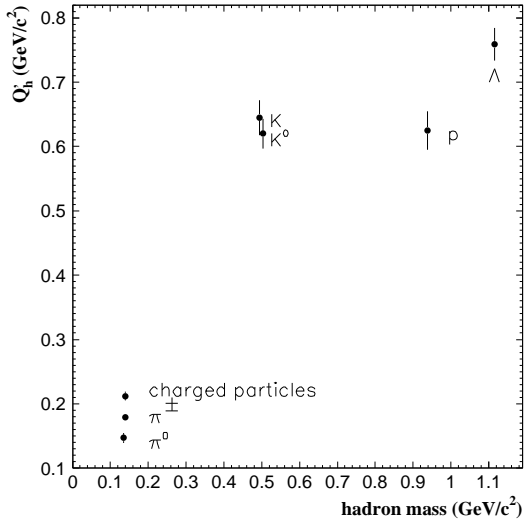,width=.9\hsize}}}
\\
\caption{The values of $\qh$ from table \ref{t:xistq2}
versus the mass of the hadron.
\label{f:q0prime}}
\end{figure}

\section{Dependence of $\xi^*_p$ on $\protect\sqrt{s}$}

A more quantitative comparison is possible for the
scaling behaviour of $\xi^*_p$, as parametrised by
the effective QCD scale $\lameff$.
The dependence on $\sqrt{s}=2\ebeam$
of the experimental values of $\xi^*_p$ from table~\ref{t:expxist}
is presented in figure \ref{f:xistq2all}.

As a start, the experimental values can be fitted
to separate first- or
second order polynomials in $\log\ebeam$ for each type of
particle. The quality of these fits provides an internal
check of the estimated experimental errors.
Table~\ref{t:xistq2} gives the slopes of fitted straight lines
and the values of $\chi^2$ for first and second
order polynomials.
The quality of the fits is acceptable, but the values of $\chi^2$
when fitting so many free parameters could indicate that the
experimental errors have been underestimated somewhat.
It seems that the measured slopes are not quite equal to one other, as
was predicted by LPHD + MLLA. This is due to some dependence
on the flavour of the particles and of the primary quarks,
violating the LPHD assumption slightly.

For a comparison with other publications it is also interesting to fit the
data for each hadron type to equation (\ref{eq:xistq2}),
under the assumption that $\qh\stackrel{\mbox{\tiny def}}{=}\lameff=Q_0$,
so that $F(\lambda)=F(0)=0$. The energy scale $\qh$ is
different from $Q_0$ as determined in the previous
section, and it is also different from $\lameff$.
In many past publications the notation `$\lameff$' is also used for $\qh$.
This is confusing, since $\qh$ is not expected to
be the same for different hadrons, while the QCD scale $\lameff$
should be independent\footnote{
In principle one might expect that different hadrons
from different decay chains could introduce
a small dependence of an effective value of $n_f$ on the
type of hadron. However, equation (\ref{eq:yuri}) shows that
variations of $n_f$ from 3 to 5 can only change the slope by a few percent.
This is well within the other uncertainties.}.
The quality of this fit with $\lameff=Q_0\equiv\qh$ is
slightly worse, with an overall $\chi^2=2.41$.
The largest contributions to $\chi^2$ come from the kaons, that also
have a different slope $b_h$ from the other hadrons. Looking carefully
at figure \ref{f:xistq2all} one might also suspect the measurements
done at the $\PZz$ and $\Upsilon$.

The fitted values of $\qh$ are not equal to
the corresponding values of $Q_0$ shown in figure~\ref{f:ximax}.
The values of $\qh$ are quite near to
the particle mass for mesons, but for the
baryons $\qh$ is significantly smaller.
Figure~\ref{f:q0prime} shows the dependence of $\qh$ on the
hadron mass. No clear linear dependence is observed.

To check how well the data is described by the calculations,
a simultaneous fit of equation (\ref{eq:xistq2}) was done
to all the data from table~\ref{t:expxist}, where
$\lameff$ is a universal scale, while $F$ is
allowed to have different values for
hadrons, pions, kaons and the two baryons.

Such a six parameter fit was performed by minimising
$\chi^2$ using the MINUIT \cite{MINUIT} program, with zero as
starting values for $F_h$. There is a strong correlation in the fit
of $\log\lameff$ and $\bar{F}$, the average of $F_h$.
Therefore it is practical to fit $\bar{F}$ and differences $F_h-\bar{F}$.
After this, the fit converges properly,
but the correlation coefficient of $\log\lameff$
and $\bar{F}$ is still $100\%$.

The strong correlations can be understood from equation (\ref{eq:xistq2}) and
figure~\ref{f:xistq2all}: a change in $\log\Lambda$ translates
the curves in the direction of the
$x$-axis, while a change of $\bar{F}$ causes a translation along the $y$-axis.
The fact that the curves are so  straight causes the strong correlations.
The reason that the fit still converges is that the slope of the
curves varies slowly with changing $\lameff$.

To see if this causes numerical instability or even inaccuracy,
it was subsequently checked that the correlation in the fit can
be reduced to a more acceptable $74\%$ by a change of variable to
$F'=\bar{F}-r\cdot(\log\Lambda-\log\Lambda_0)$,
where $r=0.62$ is the average value of $\partial\xi^*_p/\partial{Y}$ as given
by equation (\ref{eq:xistq2}) and the constant $\Lambda_0\approx\lameff$.
The fit in this parametrisation
converges to the same values of $\chi^2$ and $\lameff$
as when fitting $\bar{F}$ itself.

The fit has $47$ degrees of freedom and leads to
$\chi^2/47=2.15$.
This value of $\chi^2$ should be compared
to the values of the `total' $\chi^2/\txt{n.d.f}$
in table~\ref{t:xistq2}, that are also larger than $1$.

The resulting parameters and their statistical accuracy (after
applying the scale factor $\sqrt{\chi^2/\txt{n.d.f.}}=1.47$ to all the
experimental errors) are:
\beq
\begin{array}{lc r@{.}l@{}l}
\lameff                      &=&  0&052&{\,}_{-0.044}^{+0.12}~\GeV \\
\Longrightarrow\alpha_s(\mr{m}_{\PZz}) &=& 0&103&{\,}\pm0.022 \\
\bar{F}                      &=& -1&29 &{\,}_{-1.1}^{+0.75} \\
F_\pi         -\bar{F}       &=&  0&54 &{}\pm 0.021 \\
F_{\text{charged}}-\bar{F}   &=&  0&42 &{}\pm 0.027 \\
F_{\mr{K}}    -\bar{F}       &=& -0&28 &{}\pm 0.027 \\
F_{p}         -\bar{F}       &=& -0&28 &{}\pm 0.039 \\
F_{\Lambda}   -\bar{F}       &=& -0&41 &{}\pm 0.030 \\
\end{array}
\label{eq:fitresults}
\eeq
The errors here are purely statistical, as given by
the MINOS method in MINUIT \cite{MINUIT}, which takes into
account non-linearities and correlations between parameters.
The correlation coefficient of $\lameff$ and $\bar{F}$ is $100\%$,
while the values of $F_h-\bar{F}$ are constrained to have a zero sum.

Figure~\ref{f:xistq2all} shows that the
fitted functions are very near to straight lines and
that they follow the data points rather well.
Addition of a free $\order(Y^{3/2})$ term as in
equation (\ref{eq:xistq2}) does not improve the
fit significantly.

So what should one conclude from
the high values of $\chi^2$? It seems that the general
behaviour is described well by LPHD + MLLA, but that
equation (\ref{eq:xistq2}) differs significantly from the
data for each hadron. This is probably a signal of
the breakdown of LPHD at this level of accuracy.

An alternative way of fitting the data is based
on equation (\ref{eq:yuri}) with a fixed value of $\alpha_s$.
Allowing $\alpha_s$ to `run' somewhat leads to the following
parametrisation:
\beq
\begin{array}{l}
\xi^*_p = p_{0h} +       p_1(\log\ebeam/{\mr GeV}-\log 15) \\
  \hspace{2em} {}+ \thalf p_2(\log\ebeam/{\mr GeV}-\log 15)^2,
\end{array}
\eeq
where the quadratic term is centred at $15~\GeV$.
The constant term $p_{0h}$ is different for each hadron type.
Equations (\ref{eq:yuri}) and (\ref{eq:yuri2}) relate
the values of $p_1$ and $p_2$  to the strong coupling
constant $\alpha_s$.
This fit has no problems with correlated parameters and it converges to:
\beq
\begin{array}{l@{{}={}}r@{.}l@{}l c l@{{}={}}r@{.}l@{}l}
p_1 & 0&613&{\,}\pm0.015
 &\Rightarrow& \alpha_s &  0&122&{\,}_{-0.030}^{+0.035}
\\
p_2 & -0&012&{\,}\pm0.026
 &\Rightarrow& \alpha_s &  0&14&\pm 0.16
\end{array}
\label{eq:fitalpha}
\eeq
where $\alpha_s$ is given at a scale of $15~\GeV$.
Again, the given errors are purely statistical. The value of $\lameff$
from equation (\ref{eq:fitresults}) gives
$\alpha_s=0.123\,^{+0.033}_{-0.031}$ at the scale $15~\GeV$, which is
equal to the value derived from $p_1$.
Within its large error, $p_2$ is also consistent.
The values of $\chi^2/{\txt{n.d.f}}$ are comparable to those in the
previous fit.
The differences $F_h-\bar{F}$ can be determined from $p_{0h}$ and
they are consistent with those given in eq.~(\ref{eq:fitresults}).

\section{Dependence of $\xi^*_p$ on the hadron mass}

The dependence of $\xi^*_p$ on the hadron mass
as determined from LEP data
is shown in figure~\ref{f:ximax}(a).
As has been said, this dependence is not predicted by the LPHD + MLLA
approach, but it is interesting to look at the relation between
the scale $Q_0$ and the mass and flavour of the hadron.
Using equations (\ref{eq:xistq2}) and (\ref{eq:lambdaF})
and the determined value of $\lameff=50~\MeV$,
one can convert the values
of $\xi^*_p$ (eq.~\ref{eq:fitresults}) to values of $Q_0/\lameff$.
The result of this is shown in figure \ref{f:ximax}(b).
The error bars in this figure do not incorporate the uncertainty of
$\lameff$. This means that the absolute scale should not be taken
too seriously.

Naively one could expect that the cutoff scale $Q_0$ grows
proportionally to the mass of the hadron.
The most naive guess would be a linear dependence of the form
$Q_0\approx \Lambda_{\text{eff}}+m_{\text{hadron}}$.
Figure~\ref{f:ximax}(b)
shows that this is not correct and that at least a
separate treatment of mesons and baryons is necessary.

Figure~\ref{f:ximax}(c) also shows $Q_0$, but now
determined from the fitted $F_h$ in equation (\ref{eq:fitresults}).
Again the errors given do not include the very large overall error
on $\bar{F}$ and they are only significant when comparing the
different points relative to each other. Taking this into account,
the result is comparable to that of figure~\ref{f:ximax}(b).

\section{Summary}

The data of LEP are giving a wealth of information about the
fragmentation of quarks and gluons, as well as the mechanisms
at play in the hadronisation.
Although Monte Carlo models like {\tt JETSET} and {\tt HERWIG}
can describe these data very accurately, it is a valid question
whether these models with all their free parameters lead to
a better understanding of the physics
behind the formation of hadrons in a jet,
or are merely to a better parametrisation.

The `LPHD + MLLA' approach
\cite{book-ptqcd,mlla-xp-mp,mlla-xp-Leff,mlla-xp}
is an interesting attempt to understand some properties of
multihadron production in jets in terms of perturbative QCD.
This approach tries to take perturbative QCD as near as possible
to the limits posed by confinement.
It gives predictions for properties of momentum spectra that
can be compared with experimental data, but concentrates
on observables that can be described
without too many unknown free parameters.

The experimental values of the parameter $\xi^*_p$ for
various hadrons and at various centre of mass energies
was determined from the published momentum spectra.
Subsequently, an investigation was made
of the dependence of $\xi^*_p$ on both the
centre of mass energy and the mass of the identified hadrons,
and this could be compared with the theoretical predictions.

The dependence of $\xi^*_p$ on the mass and flavour
of the identified hadron raises some interesting questions that
have not yet been answered in the context of LPHD + MLLA and
the description of fragmentation by truncated parton cascades.

An attempt was made to clear up some of the confusion in the
definition of $\lameff$. The author would like to stress
that it is a common but confusing practice to use the
symbol $\lameff$ for the other scale, here called $\qh$.
This is because the $\qh$ is not expected to
be independent of the type of hadron, and should therefore not
be interpreted as $\Lambda_{\txt{QCD}}$.

The price paid for a
more consistent definition of $\lameff$ is the explicit introduction
of the constants $F$, that are different for different hadrons.
The advantage is that $\lameff$ can now be interpreted
as a QCD scale $\Lambda_{\txt{QCD}}$, and that it is then possible
to make a more quantitative comparison with theory.

The dependence of $\xi^*_p$ on the centre of mass energy is described
adequately by the MLLA calculations. A value of $\alpha_s$ has now
been extracted from these data, and its value is consistent with the many
accurate measurements.
It is perhaps striking that data at very low momenta
(at LEP, $p\approx 0.83 (2.27) \GeV$ when
$\xi_p\approx 3(4)$), that are
near to the region of phasespace where confinement occurs,
can not only be described qualitatively by perturbative QCD,
but can even be used to extract a
consistent value of the strong coupling constant.

It is a success of LPHD + MLLA that
the extracted value of $\alpha_s$ is correct.
However, it is clear that this is not a good way to determine
$\alpha_s$ accurately: it is nearly impossible to
make a proper estimate of the systematic errors.
The quality of the overall fit suggests that
the picture of LPHD is starting to break down.

\section*{Acknowledgements}
The author thanks  Bert Koene, Paul Kooijman and Ian Butterworth
for their critical comments.
He is grateful for helpful comments and suggestions
from Yuri L.~Dokshitzer and Valery A.~Khoze.



\end{document}